\documentclass[a4paper,twoside]{article}

\usepackage{epsfig}
\usepackage{subcaption}
\usepackage{calc}
\usepackage{amssymb}
\usepackage{amstext}
\usepackage{amsmath}
\usepackage{amsthm}
\usepackage{multicol}
\usepackage{pslatex}
\usepackage{apalike}
\usepackage{SCITEPRESS}     

\begin{document}

\title{Reduced Precision Strategies for Deep Learning:\\ A High Energy Physics Generative Adversarial Network Use Case}

\author{\authorname{Florian Rehm\sup{1,2}, \\
Sofia Vallecorsa\sup{1}, Vikram Saletore\sup{3}, Hans Pabst\sup{3}, Adel Chaibi\sup{3}, Valeriu Codreanu\sup{4}, \\Kerstin Borras\sup{2,5}, Dirk Krücker\sup{5}}
\affiliation{\sup{1}CERN, Switzerland}
\affiliation{\sup{2}RWTH Aachen University, Germany} 
\affiliation{\sup{3}Intel, USA}
\affiliation{\sup{4}SURFsara, Netherlands}
\affiliation{\sup{5}DESY, Germany}

}


\abstract{
Deep learning is finding its way into high energy physics by replacing traditional Monte Carlo simulations. However, deep learning still requires an excessive amount of computational resources. 
A promising approach to make deep learning more efficient is to quantize the parameters of the neural networks to reduced precision. 
Reduced precision computing is extensively used in modern deep learning and results to lower execution inference time, smaller memory footprint and less memory bandwidth.
In this paper we analyse the effects of low precision inference on a complex deep generative adversarial network model. The use case which we are addressing is calorimeter detector simulations of subatomic particle interactions in accelerator based high energy physics. 
We employ the novel Intel low precision optimization tool (iLoT) for quantization and compare the results to the quantized model from TensorFlow Lite.
In the performance benchmark we gain a speed-up of 1.73x on Intel hardware for the quantized iLoT model compared to the initial, not quantized, model. With different physics-inspired self-developed metrics, we validate that the quantized iLoT model shows a lower loss of physical accuracy in comparison to the TensorFlow Lite model.
}

\onecolumn \maketitle \normalsize \setcounter{footnote}{0} \vfill

\section{\uppercase{Introduction}}
\label{sec:introduction}

Simulations in High Energy Physics (HEP) have, due to their complex physical processes, a high demand for computational hardware resources and require more than half of the worldwide Large Hadron Collider (LHC) grid resources \cite{RoadmapHEP}. 
Currently, the interactions of particles in the detectors are simulated by employing the simulation toolkit Geant4 \cite{Geant4} which uses detailed Monte Carlo simulation for reproducing the expected detector output. Because of their dense material and the highly granular geometry, calorimeters are among the detectors that take longer time to simulate.
Future simulations of the LHC in the high luminosity phase will require around 100 times more simulated data which exceeds drastically the expected computation resources, even taking into account technological development \cite{HL-LHC}. Therefore, intense research is already ongoing for seeking faster alternatives to the standard Monte Carlo approach.

Several prototypes based on Deep Generative Models have shown great potential, by achieving similar accuracy than traditional simulation \cite{de2017learning,Salamani2018}. Generative Adversarial Networks (GANs) represent an example of such model that have proven for creating calorimeter shower images \cite{EnergyGAN}.

In this paper we describe our developments to further enhance the GAN approach for calorimeter simulations by quantizing the trained neural network into lower precision. Lower precision operations lead to a reduced computation time, smaller model storage requirements and fewer data movements. We assess the impact of quantization on the computational side by measuring the inference time and the GPU memory footprint. On the physical side we validate the accuracy by comparing the energy deposition patterns ("shower shapes") observed in GAN and Geant4 images. We benchmark the new Intel Low Precision Optimization Toolkit \cite{iLoT} by quantizing our model using Integer-8 (int8) format and compare the results to the quantized models obtained using TensorFlow Light \cite{TFLite} in 16-bit floating point (float16) and int8 formats.

Section 2 provides a description of calorimeter simulations and the data set format. Section 3 briefly introduces the GAN approach and the prototype we use for this study. Section 4 covers an introduction to reduced precision and to the quantization tools. A brief review on related work is shown in section 5. In section 6 results are evaluated in terms of computational resources and physical accuracy. The last section summarizes the conclusions.

\section{\uppercase{Electromagnetic Calorimeters}} \label{sec:Calorimeters}

\noindent 
Calorimeters are one of the main components of accelerator-based HEP experiments and detectors. They are used to measure the energy of particles \cite{ECAL} at collider experiments. The primary particles which enter and interact with the material of the calorimeter deposit their energies by creating showers of secondary particles as they pass through the detector. The secondary particles in turn, create other particle showers by the same mechanism. While the shower evolves, the energy of the particles gradually reduces and is absorbed or measured by the calorimeter sensors. 
In our work, we simulate electromagnetic calorimeters, which are specially designed to measure energies of electrons, positrons and photons that interact in the detector volume via electromagnetic interactions (mainly bremsstrahlung for electrons and pair production for photons).
\newline

\begin{figure}[]  
    \centering
    \includegraphics[width=.49\textwidth]{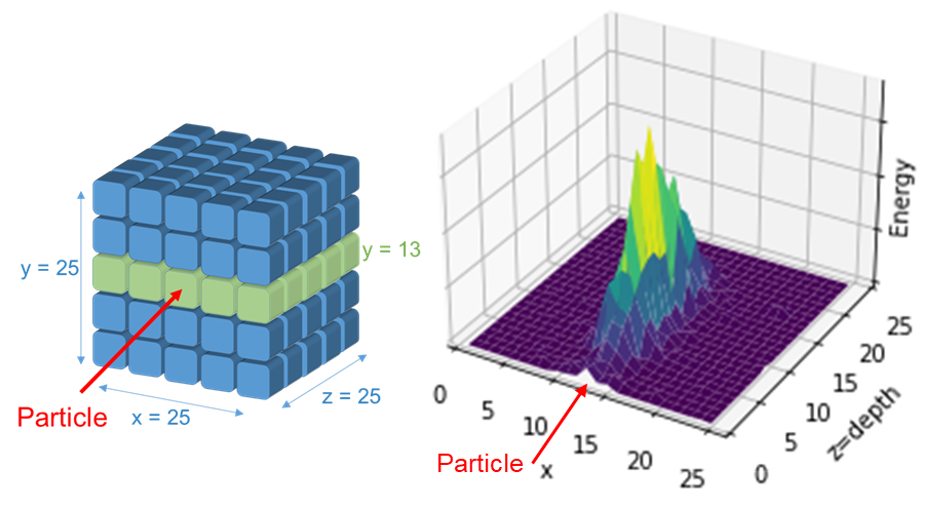}
    \caption{(left): A simplified representation of the 3-dimensional image. (right): An example shower development at $y=13$.}
    \label{fig:example_shower}
\end{figure}

We use particle shower images, recorded by the calorimeter, as data set for the simulations. The training and test data set are simulated using Geant4: they represent a future electromagnetic calorimeter prototype. For the study in this paper, we are using 200 000 3-dimensional shower images of electrons with a primary particle energy in the range of 100-500 GeV and a 3-dimensional shape of 25x25x25 pixels. Figure \ref{fig:example_shower} shows an example shower image cutout.

It is worth noticing that we do not use any timing information: only static images of the energy deposits are used to train the GAN model (described in section \ref{sec:GAN}). Particle shower shapes are parameterized according to the energy of the primary particle $E_p$, which is the energy of the electron which enters the calorimeter volume and the ECAL value (for Electromagnetic CALorimeter), which corresponds to total energy measured by the calorimeter (summing the estimated energy deposits over all calorimeter cells). The relation between ECAL and  $E_p$  depends on the type, geometry and material of the calorimeter. For our data we approximate it using polynomial function.

\section{\uppercase{3D Generative Adversarial Networks}}  \label{sec:GAN}
\noindent 
Generative Adversarial Networks (GANs) are being successfully investigated for replacing some traditional Monte Carlo calorimeter simulations. It is demonstrated, that GANs can achieve similar levels of accuracy as the Monte Carlo simulations, on some specific use case, while considerably decreasing the simulation time \cite{EnergyGAN}  \cite{WGAN_Thorben}. 
GANs were first introduced by Ian Goodfellow \cite{goodfellow} in 2014. They belong to the group of unsupervised learning methods and are nowadays used for a large variety of different generative tasks. The whole GAN model consists of two deep neural networks. A generator network which generates images from random numbers and a discriminator which is trained to evaluate and distinguish between the generated and the training images. 

\begin{figure*}[ht!]  
    \centering
    \includegraphics[width=.97\textwidth, clip=true]{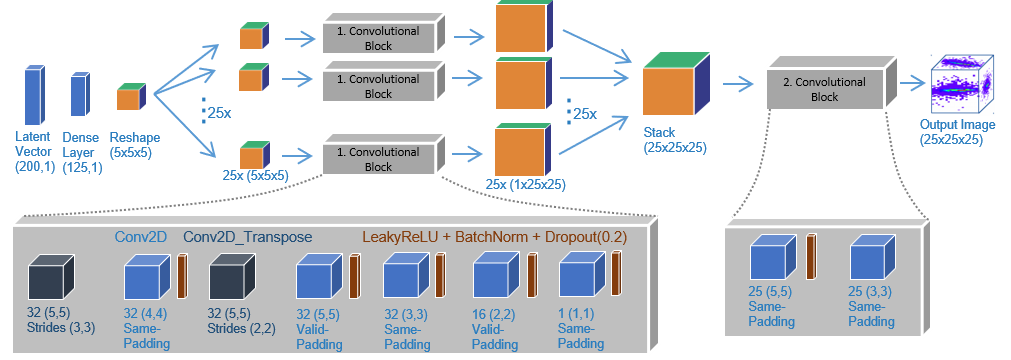}
    \caption{Generator neural network architecture. Input is a latent vector and output is the 3-dimensional shower image.}
    \label{fig:generator}
\end{figure*}

The generator network receives as input a latent vector with 200 uniform distributed random numbers multiplied by an energy scalar corresponding to the primary particle energy $E_p$. It produces a 3-dimensional shower image with the same shape 25x25x25, as the images of the training dataset. The generator network uses a mix of convolutional 2D (Conv2D) layers and transposed convolutional 2D (Conv2D\_transpose) layers. Additionally, we use batch normalization (BatchNorm), leaky rectified linear units activation function (LeakyReLU) and dropout layers (Dropout). The detailed neural network architecture of the generator is shown in figure \ref{fig:generator}.

\begin{figure*}[ht!]  
    \centering
    \includegraphics[width=.97\textwidth, clip=true]{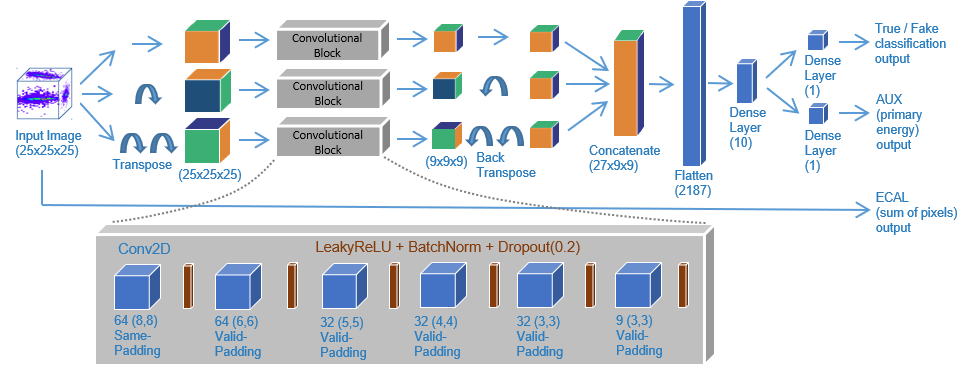}
    \caption{Discriminator neural network architecture. Input are the 3-dimensional shower images and output the loss values.}
    \label{fig:discriminator}
\end{figure*}

The discriminator produces three output values, see figure \ref{fig:discriminator}. The first is the typical GAN true/fake probability \cite{goodfellow}. The second and third discriminator outputs are auxiliary losses to support the GAN model to converge during training. We name the second loss AUX (for AUXiliary loss) which is a regression task to compare the discriminator output with the label of the primary energy $E_p$. The third discriminator output is named ECAL and is a self-produced lambda layer calculating the sum over the pixels of the input image which therefore, corresponds to the total energy of the input image. 

In previous papers we used convolutional 3D (Conv3D) layers in the models but here we replace them by Conv2D layers. The Conv2D layers reduce the computational complexity with respect to the Conv3D approach and we gained a reduction in computation time. In addition the Conv3D layers are not supported in any quantization tool yet. We will describe the change from Conv3D to Conv2D more detailed in a future paper, since this approach is applicable for any kind of convolutional models with 3D images and comes with a significant decrease in computation time.

\section{\uppercase{Reduced Precision Computation}}  \label{sec:ReducedPrecisionComputation}
\noindent 
To achieve higher throughput and to reduce memory bandwidth during inference we can shrink the size of activations and weights of the trained neural network to lower precision. This approach is named reduced precision computing and the process of converting numbers from higher to lower precision is named quantization. The standard number format in machine learning is floating point 32 (float32) or single precision, specified in IEEE 754 \cite{float32}. 
The format which uses just half of the bits of float32 is floating point 16 (float16) and is therefore called as half precision, specified in IEEE 754 \cite{float32}. 
The smallest format in which we want to quantize our model is integer-8 (int8). 
Integer numbers can be either signed int8 (sint8) with a range of $[-127,127]$ or unsigned int8 (uint8) with a range of $[0,255]$.\\

\textbf{Intel Low Precision Optimization Tool} \\
The Intel Low Precision Optimization Tool (iLoT) \cite{iLoT} is an open-source python library for quantizing deep learning models and running low precision inference across multiple frameworks. It uses the Intel oneAPI Deep Neural Network Library (oneDNNL) \cite{oneDNNL}, which contains building blocks for deep learning applications to improve the performance on Intel processors. This paper presents results of the first test of the brand new iLoT tool on a real use case.

Quantization of a trained model requires identifying the best order of magnitude across pixels covering an entire layer instead of a single pixel. Once the order of magnitude of an entire layer is known, it is possible to drop single bits, named "outliers", which are far away from the other values within this layer. Dropping outliers, the choice of different statistical means and repeated refinement of the topology makes quantization a delicate process rather than just a one-step deployment task. 
iLoT tunes the quantization process automatically with accuracy-driven strategies for computational performance, model size and memory footprint \cite{iLoT}. During the iterative and automatic quantization process iLoT can keep a few single nodes in the neural network in float32 precision in order to increase the accuracy. This is the advantage of iLoT compared to any other recent quantization tool. The accuracy is measured using a physics-inspired validation metric introduced in section \ref{sec:evaluation}.

Previously, quantization in deep learning was mainly applied to classification problems which employ simpler neural network outputs (probability values of classes) compared to the complex shower images of our network. Moreover, our generator network uses LeakyReLU activation functions instead of standard ReLU. Since LeakyReLU can generate negative weights, they require sint8 operations after quantization. As the sint8 operations are not needed for classification networks with standard ReLU activations, sint8 are by default not yet implemented in TensorFlow. Similarly no recent quantization tool implements LeakyReLU activations functions. In this study we make use of an Intel-customized TensorFlow version supporting sint8 operations and an ad-hoc implementation of LeakyReLU activations in the iLoT tool.

Layers fusion is another strategy that can be used to accelerate deep neural networks inference: iLoT implements [Conv2D + LeakyReLU] fused operation followed by a FusedBatchNorm layer. In its next release, iLoT (included in oneDNNL) will support a full [Conv2D + LeakyReLU + BatchNorm] fuse, which could further improve the performance.
\newline

\textbf{TensorFlow Lite} \\
TensorFlow Lite \cite{TFLite} is part of the TensorFlow library and is created for deploying machine learning models on mobile and internet of things (IoT) devices. It contains packages for converting TensorFlow models to TensorFlow Lite models and packages to quantize and deploy models. It can quantize networks, layer-wise, from float32 into float16 and int8 \cite{TFLite}. 

Because TensorFlow does not support sint8 operations yet, TensorFlow Lite does not support quantized LeakyReLU activation functions as well. Additionally, we do not have the compatible hardware for running quantized TensorFlow Lite models available for this study, therefore we use TensorFlow Lite only for measuring the accuracy of the quantized models and not inference speed-up.

\section{\uppercase{Related Work}}

\textbf{Generative Adversarial Networks} \\
GANs are today extensively used in a wide range of applications \cite{GANexamples}. 

In the field of High Energy Physics, a lot of research is ongoing to understand how to replace Monte Carlo simulations by employing GANs. For example in \cite{GANandVAE} Variational Auto-Encoders (VAEs) and GANs are proposed to simulate the output of the ATLAS liquid Argon calorimeter.
The first proof of concept, using 3D convolutional GANs for electromagnetic calorimeters is represented by the 3DGAN prototype \cite{EnergyGAN} and \cite{angleGAN} based on a simple auxiliary GAN model. Other approaches investigated Wasserstein GAN \cite{WGAN} for improving the model convergence \cite{WGAN_Thorben}.
\newline

\textbf{Int8 Quantization} \\
Quantizing trained deep learning models to a reduced precision is being actively researched in the recent years to allow, in particular, fast inference on mobile devices where the computational power is strongly constrained. In more general, with increasing model size and complexity, constraints on computing time and memory consumption become relevant for other architectures, CPUs and accelerators alike.

In most cases, quantization techniques are targeted to int8, because it is a data format that most modern hardware supports and because it maintains almost the same level of accuracy, compared to lower precision formats \cite{Quantization}. Most of the existing benchmarks represent classification problems \cite{classification_quantization}, while in our case, we require an accurate description of simulated data. There are two main quantization techniques. The first is post-training quantization, where the model is trained in float32. Afterwards, it uses a calibration dataset to calculate the maxima of the weights and activations which are needed for the quantization process into a lower precision. This is commonly the first approach which is chosen and leads mostly to a satisfying level of accuracy \cite{nvidia_quantization}, hence we use this approach for the research in this paper. An alternative method, named quantization-aware training, directly trains models using lower precision formats.
\newline

\textbf{Mixed precision training}\\
One can go even further and train models with mixed precision formats: for example, using both float32 and int8. In this case, some weights are represented in int8 format to gain a speed-up, but others are kept as float32, to maintain the level of accuracy. Choosing which weights are represented in higher or lower precision is done by intelligent algorithms.

Different studies have evaluated  mixed precision for training \cite{MixedPrecision} \cite{MixedPrecision2}: In particular \cite{IntelMixedPrecision} applies a mixed precision approach to the training of 3DGANs.

\section{\uppercase{Evaluation}} \label{sec:evaluation}
\noindent 
This section describes the evaluation process in terms of computational performance and physics accuracy and discusses results in details.

\subsection{Computational Evaluation} 
\noindent 
The computational throughput depends in practice on the model and the hardware on which it runs. We perform all computational performance tests on an Intel 2S Xeon Processor 8280 with Cascade Lake architecture and 28 cores (56 virtual cores or threads) which supports int8 format. As explained above, the iLoT int8 model is benchmarked against the original float32 model.

\begin{figure}[ht!]
    \centering
    \includegraphics[width=.47\textwidth]{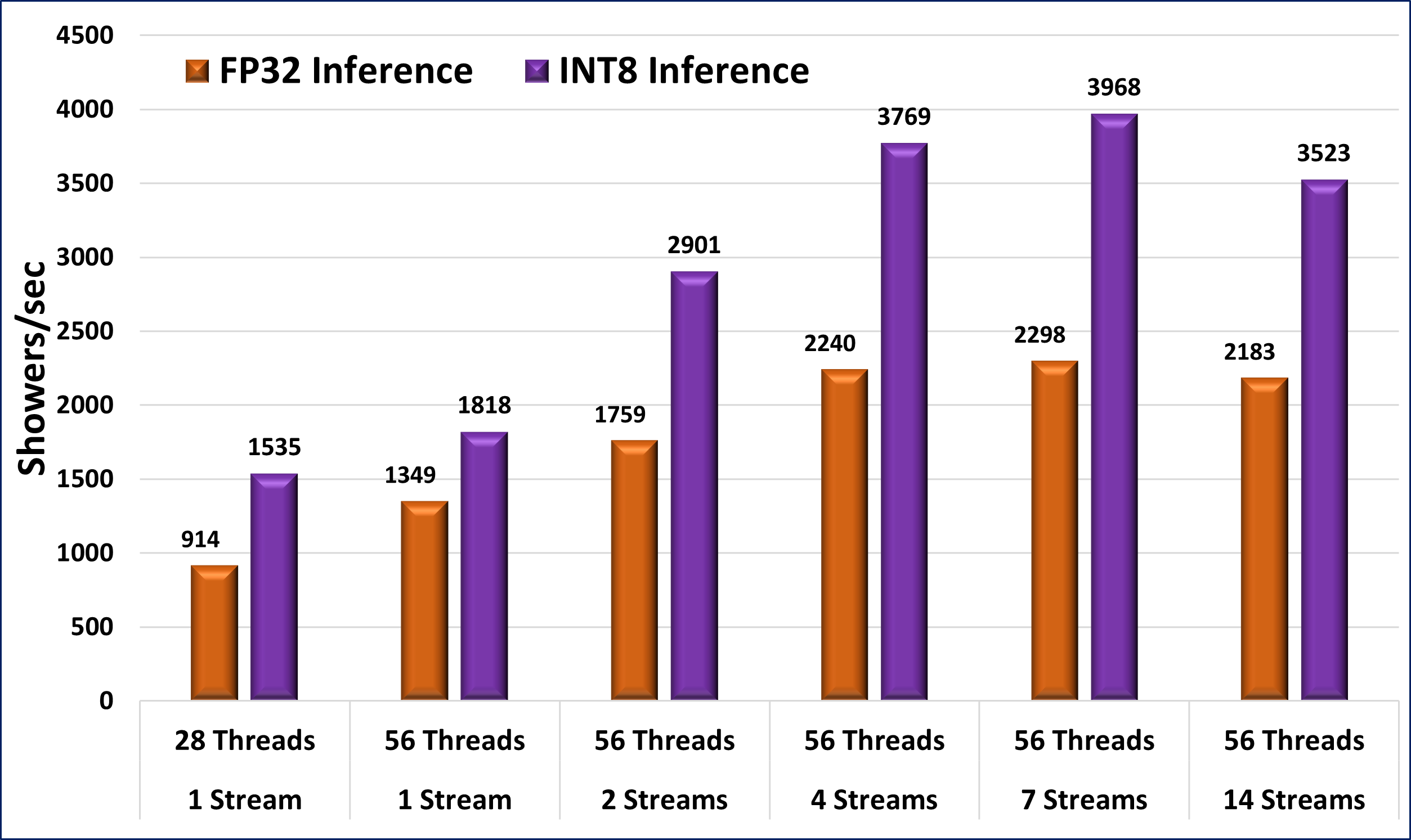}
    \caption{Inference throughput in showers per seconds for the float32 and int8 model for different number of stream and thread configurations.}
    \label{fig:SpeedUp}
\end{figure}

Multiple flows of different input data are processed concurrently on affinitized threads (and taking advantage of memory locality), which is referred to multiple streams. We measured the total time and the throughput by parallelising the inference process using multiple threads and multiple data streams: 
The results are shown in figure \ref{fig:SpeedUp}. As expected, the number of showers per seconds increases with the number of streams and threads for both the int8 and float32 models. In particular the best performance is achieved using 56 threads and 7 data streams. Beyond this point the performance becomes worse, most probably because the memory bandwidth limit is exceeded and the CPU gets oversubscribed. All tests were run including a warm-up time with TensorFlow version 2.3.

Using 7 streams and 56 threads for both models we measure an inference time of 55.7 ms for the float32 model and 32.3 ms for the int8 model, as shown in table \ref{tab:speedup}, corresponding to a 1.73x speed-up using the iLoT quantized model. 
 
\begin{table}[h!]
 \centering
 \caption{Inference times in milliseconds [ms] for float32 and INT8 GAN model measured on Cascade Lake CPU. }
 \begin{tabular}{|c c c|} 
 \hline
 float32 & int8 & Speed-up \\ [0.5ex] 
 \hline
 55.7 ms &	32.3 ms &	1.73x \\ 
 \hline
\end{tabular}
 \label{tab:speedup}
\end{table}

Details about the runtimes of some operations are shown in table \ref{tab:runtimes}. It can be seen, that the FusedBatchNorm yields about 30\% of the total runtime: replacing this operation with a fused [Conv2D + LeakyReLU + BatchNorm] will bring a significant speed-up.  

\begin{table}[h!]
 \centering
 \caption{Times in milliseconds (ms) for some operations of the quantized int8 iLoT model.} 
\begin{tabular}[h]{|c c c|} 
 \hline
  Total & FusedBatchNorm &  De-/Quantization \\ [0.5ex] 
 \hline
 32.3 ms &  	9.37 ms &		6.06 ms \\ 
 \hline
\end{tabular}
 \label{tab:runtimes}
\end{table}

Table \ref{tab:runtimes} also shows the time needed by the de-quantize operation, which is non negligible as it corresponds to $\frac{6.06 \:ms}{32.3 \:ms}=  18.3\%$ of the total runtime. The dequantization step takes care of converting the float32 input data to int8 and then converting back the model output data (int8) to float32, expected for the simulation output.

An additional effect is due to the fact that, a low-precision unit internally demands higher bandwidth, i.e. scratch memory bandwidth or cache bandwidth, and this demand can limit operations depending on the specific hardware implementation. Therefore, practical implementations of low-precision in hardware and software are challenged to speed-up inference by the full ratio of type-width. 

Generally speaking, larger models might exhibit a larger inference acceleration, simply thanks to the relatively smaller weight represented by the quantization/dequantization steps with respect to the quantized operations runtime. As an example, ResNet-50 \cite{resnet50} has over 23 million parameters, Inception-v3 \cite{inceptionv3} over 24 million parameters whereas our model has only around 2 million parameters. Being 10 times smaller, our GAN network is more sensitive to the size of the quantization and dequantization steps (10.5 ms) with respect to the total run time (32.3 ms). In other words, the smaller the model, the stronger the impact of the quantization and dequantization functions at the beginning and end of the model.

In any case, running inference using int8 precision, further increases the advantage with respect to running standard Monte Carlo simulation. With respect to the original Geant4 benchmarks quoted in \cite{angleGAN}, the int8 quantized models reaches a huge 67 000x speed-up. 
In terms of memory consumption, the iLoT int8 model decreases by a factor 2.26x the total utilised physical memory (from 8 KB to 3.6 KB) with respect to the initial float32.

\subsection{Physical Evaluation} 
\noindent 
Evaluating performance of a Generative Model is not an easy task, several methods have been proposed depending on the specific applications  \cite{GANevaluation}. 
In this work, we rely on physics inspired validation, visually inspecting the energy patterns across the calorimeter volume by measuring the accuracy on specific physics quantities, the energy deposits distributions along the calorimeter $y$- and $z$-axis. 

We compare the quantised models in three different formats (int8 iLoT, int8 TFLite and float16 TFLite) to the baseline GAN model (float32) and the Geant4 validation set.
\newline

\begin{figure}[ht!]
    \centering
    \includegraphics[width=.45\textwidth]{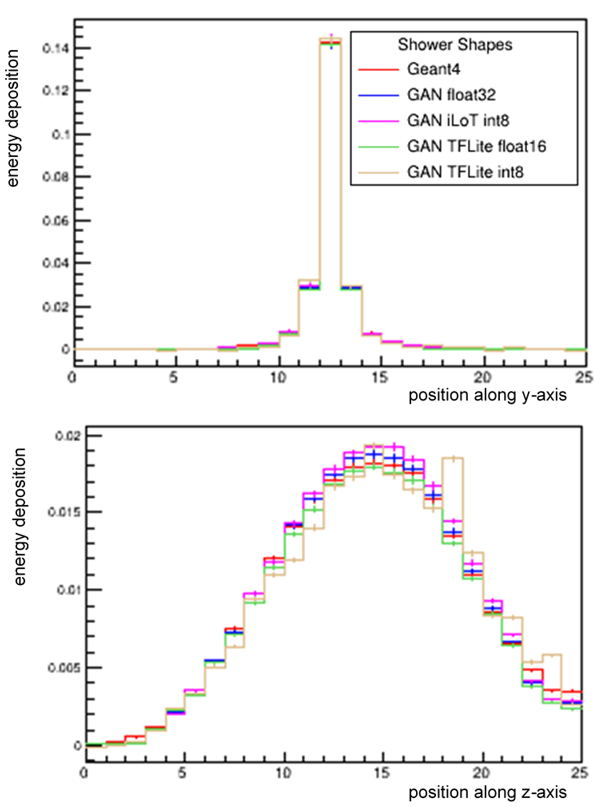}
    \caption{Generated shower shapes along y- (top) and z-axis (bottom) of Geant4 (red), the TF float32 model (blue), the iLoT int8 model (pink), the TFLite float16 model (green), and the TFLite int8 model (orange). Horizontal we plot the single pixels of the respective axis and vertical the average energies contained in the single pixels. The vertical lines corresponds to the statistical errors for the single pixels.}
    \label{fig:hist0}
\end{figure}

\textbf{1. Particle Shower Shapes} \\
Figure \ref{fig:hist0} represents the first visual validation method and shows the 2-dimensional particle shower shapes along the $y$- and the $z$-axis for the different models. The particle enters the detector at the coordinates $x=13$, $y=13$ and $z=0$, orthogonally to its surface: the larger energy depositions, in the transverse ($x, y$) plane, cluster around the middle of the image while the patterns along the $z$-axis develops as discussed in section \ref{sec:Calorimeters}. For simplicity, we report only the energy distribution along the $y$-axis, since the patterns along the two transverse directions are very similar. In figure \ref{fig:hist0} we compare all models to the Geant4 prediction.  

The float32 prediction (our baseline) is very close to the Geant4 test data. The quantized iLoT int8 and TFLite float16 models are close to initial float32 model and remain mostly within the range of the statistical error bars. However, the TFLite int8 model is slightly off. This effect is larger at the edges of the distribution, where the energy deposits are much smaller (see figure \ref{fig:hist0}).

Table \ref{tab:statistics} summarizes the statistical mean and standard deviation for the particle shower distributions in figure \ref{fig:hist0}. The TFLite int8 seems to yield worse prediction with respect to both the GAN baseline (float32) and Geant4. 

\begin{table}[h!]
 \centering
 \caption{Statistical values mean and standard deviation (STD) for the different models.} 
\begin{tabular}[h]{|c | c | c|} 
 \hline
  Model: & Mean: & STD:  \\ [0.5ex] 
 \hline
 $y$-axis Geant4             & 12.00	    & 1.45	  \\
 $y$-axis float32            & 12.02	    & 1.46	  \\ 
 $y$-axis iLoT int8          & 12.03	    & 1.45	  \\ 
 $y$-axis TFLite float16     & 12.05	    & 1.13	  \\ 
 $y$-axis TFLite int8        & 12.10	    & 1.19	  \\ 
 \hline
 $z$-axis Geant4             & 13.85	    & 4.62	  \\
 $z$-axis float32            & 13.81	    & 4.50	  \\ 
 $z$-axis iLoT int8          & 13.91	    & 4.48	  \\ 
 $z$-axis TFLite float16     & 13.78	    & 4.50	  \\ 
 $z$-axis TFLite int8        & 14.25	    & 4.62	  \\ 
 \hline
\end{tabular}
 \label{tab:statistics}
\end{table}

\begin{figure}[ht!]
    \centering
    \includegraphics[width=.40\textwidth]{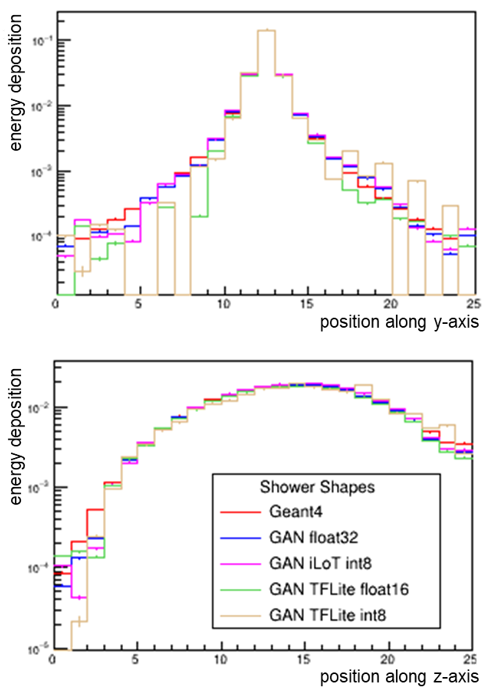}
    \caption{Logarithmic Shower Shapes.}
    \label{fig:hist_log0}
\end{figure}

This can be visually confirmed by looking at the same shower shapes in logarithmic scale in figure \ref{fig:hist_log0}. In the $y$-axis plot we can see that TFLite float16, and especially TFLite int8 struggles with the low energies at the tails. Tails we name the first and last five pixels of the y-axis shower shape plot.
On the other hand, the iLoT int8 model has no troubles in representing the low energy pixels at the tails. However, a more detailed performance analysis is needed to confirm this result. \newline 

\textbf{2. Single Validation Number} \\
In order to simplify results evaluation, we define a composite accuracy value, calculated by building 2-dimensional projections of the particle shower distributions for the GAN and Geant4 samples and then measuring the total mean squared error (MSE) between the two projections. 
The MSE-based metric values for the various models are shown in table \ref{tab:validation}. \newline

\begin{table}[h!]
 \centering
 \caption{MSE-based values for the different models. The lower the value, the better the accuracy.} 
\begin{tabular}[h]{|c c c c|} 

 \hline
  float32 & iLoT & TFLite float16 & TFLite int8 \\ [0.5ex] 
 \hline
 0.061 &	0.053 &	0.254 &	0.340 \\ 
 \hline
\end{tabular}
\label{tab:validation}
\end{table}

The iLoT model seem to reach a better accuracy than the initial float32 model. However, this is an artifact of the quantization process that relies on the same metric to determine which weights are kept in float32 format in a similar fashion as an architecture hyperparameter search. In contrast, both quantized TFLite models loose accuracy.
\newline

\textbf{3. Pixel-Wise Image Comparison} \\
In addition we perform a pixel-wise image comparison, by fixing the input latent vector and  generating a synthetic image. We  subtract then element-wise the entries of the single pixels of the output image of the quantized model from the baseline model and sum up the absolute values. The pseudo code for the pixel-wise comparison looks like: 
\begin{small}
\begin{verbatim}
  mean ( sum ( abs ( X_float32 - X_int8 ) ) )
\end{verbatim}
\end{small}

With the corresponding values we can directly compare how different the created images are. 
The smaller the pixel-wise validation value, the closer are the images and the better the model. We performed the pixel-wise validation for all our quantized models compared to the baseline float32 model. The results are shown in table \ref{tab:pixelwise}. One can see that the TFLite float16 model performs best and the iLoT int8 model is approximately three times better than the TFLite int8 model in the pixel-wise comparison.

\begin{table}[h!]
 \centering
 \caption{Pixel-wise validation values for the different quantized models.} 
\begin{tabular}[h]{|c | c | c |} 
 \hline
  Model:   &   Mean: & STD: \\ [0.5ex] 
 \hline
 TFLite float16 & 0.133 & 0.291\\ 
 \hline
 TFLite int8 & 4.054 & 0.721\\ 
 \hline
 iLoT int8 & 1.550 & 0.191\\ 
 \hline
\end{tabular}
 \label{tab:pixelwise}
\end{table}

All in all the iLoT tool reaches a better level of physics accuracy in all our shown validation metrics compared to TFLite int8.

\section{\uppercase{Conclusion and Future Work}}
\noindent 
In this paper we studied the impact of int8 quantization on a convolutional GAN model developed for High Energy Physics detector simulation. Preliminary results seem to suggest that the iLoT tool performs better than TFLite, in terms of physics accuracy. Additional studies are on-going in order to confirm these findings. We are interested, in particular, in understanding what is the effect of the optimisation metric used by the iLoT tool on the overall output quality.

Thanks to the quantization we obtain a 1.73x speed-up on inference time compared to the initial float32 model. This brings the total speed-up with respect to the standard Monte Carlo approach to several orders of magnitude (67 000x). These results make the proposed reduced precision strategies on GAN models an attractive approach for future researches on the detector simulations and could help to save large computing resources in view of the future High Luminosity LHC runs.

With this paper we hope to inspire colleagues to follow us and to apply quantization to deep learning models with more complex outputs than classification tasks. More use cases, in turn, will lead to an improvement of quantization methods and tools leading to possible further decrease of inference times in the future.

\section*{\uppercase{Acknowledgements}}

\noindent 
This work has been sponsored by the Wolfgang Gentner Programme of the German Federal Ministry of Education and Research.

\bibliographystyle{apalike}
{\small
\bibliography{ICPRAM}}

\end{document}